\documentclass{ws-mpla}

\begin{document}

\markboth{Cosimo Bambi}
{Testing the Kerr BH Hypothesis}

\catchline{}{}{}{}{}

\title{TESTING THE KERR BLACK HOLE HYPOTHESIS}

\author{\footnotesize COSIMO BAMBI}

\address{Arnold Sommerfeld Center for Theoretical Physics\\
Ludwig-Maximilians-Universit\"at M\"unchen\\
Theresienstr. 37, 80333 Munich, Germany\\
Cosimo.Bambi@physik.uni-muenchen.de}

\maketitle


\begin{abstract}
It is thought that the final product of the gravitational collapse is 
a Kerr black hole and astronomers have discovered several
good astrophysical candidates. While there is some indirect
evidence suggesting that the latter have an event horizon, and
therefore that they are black holes, a proof that the space-time 
around these objects is described by the Kerr geometry is still 
lacking. Recently, there has been an increasing interest in the 
possibility of testing the Kerr black hole hypothesis with present 
and future experiments. In this paper, I briefly review the state of 
the art of the field, focussing on some recent results and work in 
progress.
\keywords{Black holes; Modified theories of gravity.}
\end{abstract}

\ccode{PACS Nos.: 97.60.Lf, 04.50.Kd}

\section{Introduction \label{s-1}}	

General Relativity (GR) is our current theory of gravity and so far 
there are no clear observational evidences in disagreement with 
its predictions~\cite{will}. However, the theory has been tested 
only in certain regimes; in particular, only for weak gravitational 
fields. Even in the famous observation of the decay of the orbital 
period of the binary system PSR1913+16, due to the emission 
of gravitational waves, the gravitational potential is $|\phi| \sim M/r 
\sim 10^{-6}$; that is, $|\phi| \ll 1$\footnote{Throughout the paper, 
I use units in which $G_{\rm N}=c=1$.}. On the other hand, we 
know GR breaks down in some extreme situations: the theory 
allows for the existence of space-time singularities, where 
predictability is lost, and regions with closed time-like curves, 
where causality can be violated. GR seems also to be incompatible 
with quantum mechanics, and probably for this reason we do 
not have yet a quantum theory of gravity.

One of the most intriguing predictions of GR is that the collapsing
matter produces singularities in the space-time, at least in the sense 
of time-like or null geodesic incompleteness\footnote{This 
conclusion requires the following assumptions: $i)$ the validity of 
the Einstein's equations with non-positive cosmological constant, 
$ii)$ the energy condition $T^\sigma_\sigma \ge 2 T_{\mu\nu}t^\mu 
t^\nu$ for any vector field $t^\mu$ such that $t^\nu t_\nu=-1$, where 
$T_{\mu\nu}$ is the matter energy-momentum tensor and the metric 
has signature $(-+++)$, $iii)$ the absence of closed time-like curves, 
and $iv)$ every time-like or null geodesic enters a region where the 
curvature is not specially aligned with the geodesic. The condition 
$ii)$ means that gravity must be always attractive. The condition 
$iv)$ just excludes very pathological situations. For more details,
see Ref.~\cite{sing}.}~\cite{sing}. There is no theorem restricting 
the nature of these singularities, but there are only two options: the 
singularities are hidden behind an event horizon, and the final product 
of the collapse is one or more black holes (BHs), or the singularities are naked. 
Since space-times with naked singularities have several kinds of 
pathologies, one assumes the weak cosmic censorship conjecture, 
according to which all the singularities of gravitational collapse must
be hidden within BHs~\cite{wccc}. Numerical simulations support this 
conjecture~\cite{wccc2,wccc3}. Surprisingly, it turns out that in 4-dimensional 
GR there is only one uncharged BH solution, the Kerr metric, which is
completely specified by two parameters, the mass, $M$, and the spin 
angular momentum, $J$, of the object. This is the celebrated ``no-hair''
theorem~\cite{hair,hair2}, although, strictly speaking, a Kerr BH has ``two 
hairs'', $M$ and $J$. The condition for the existence of the event 
horizon is $|a_*| \le 1$, where $a_* = J/M^2$ is the (dimensionless) 
spin parameter. For $|a_*| > 1$, the Kerr metric does not describe a 
BH but a naked singularity, which is forbidden by the weak cosmic 
censorship conjecture. Astrophysical BHs presumably form from the 
gravitational collapse of matter (stars or clouds), while the no-hair 
theorem demands that the space-time is time independent. However, 
one can see that any deviation from the Kerr solution is quickly 
radiated through the emission of gravitational waves~\cite{price,price2}
(but see Ref.~\cite{mck}). Numerical simulations show that a BH 
rapidly goes ``bald'' even when the initial deviations from the Kerr
background are large (see e.g.~\cite{alcu}). Let us also notice that the 
Kerr metric is not an exact solution of the Einstein's equations only, 
but it is common to many other theories of gravity; however, in general, 
there may not be a uniqueness theorem as in 
GR~\cite{kerr}\footnote{Nevertheless, 
gravitational waves emitted by a Kerr BH in GR and by a Kerr BH in 
another theory of gravity may be different and may thus be used 
to distinguish the two scenarios. The gravitational radiation depends 
indeed not only on the metric of the space-time, but also on the field 
equations of the gravity theory~\cite{kerr2}.}.

At the observational level, there are at least two classes of
astrophysical BH candidates~\cite{ram}: stellar-mass objects in
X-ray binary systems (mass $M \approx 5-20$~$M_\odot$) and 
super-massive objects in galactic nuclei ($M \sim 10^5 - 
10^9$~$M_\odot$). The existence of a third class of objects,
intermediate-mass BHs with  $M \sim 10^2 - 10^4$~$M_\odot$, 
is still controversial, because their detections are indirect and 
definitive dynamical measurements of their masses are still 
lacking. All the BH candidates are supposed to be Kerr BHs because 
they cannot be explained otherwise without introducing new 
physics. The stellar-mass objects in X-ray binary systems are 
too heavy to be neutron or quark stars for any reasonable matter 
equation of state~\cite{kal}. At least some of the super-massive 
objects in galactic nuclei are too massive, compact, and old 
to be clusters of non-luminous bodies~\cite{maoz}. There is
also observational evidence that the surface of the BH candidates
does not emit any radiation, even when it is hit by the accreting gas. 
This fact has been interpreted as an indirect proof of the existence 
of an event horizon~\cite{horizon,horizon2} (see however Ref.~\cite{abra} 
for different interpretations). The existence of an event horizon is
also theoretically required for the stability of any rotating very 
compact object~\cite{ergo1}. However, for systems close to the 
formation of an event horizon, the instability time scale would be
very long and they could still be consistent with observations~\cite{ergo2}.
Examples of compact objects with a ``quasi-event horizon'' can be found,
for instance, in~\cite{noev1,noev2}.

The aim of this paper is to provide a short review on the current
attempts to test the Kerr nature of the astrophysical BH candidates,
focussing the attention on some selected recent results. The content of 
the manuscript is as follows. In Sec.~\ref{s-2}, I review the basic 
idea to test the nature of the BH candidates and
the theoretical frameworks proposed in the literature. In Sec.~\ref{s-3}, 
I show that non-Kerr BHs may have some fundamental properties
that are different from the ones expected for a BH in GR. In 
Sec.~\ref{s-4}, I review the possibilities offered by present and 
future experiments to test the Kerr black hole hypothesis. The
conclusions are reported in Sec.~\ref{s-5}.


\section{Measuring deviations from the Kerr geometry \label{s-2}}

A framework within which to test the Kerr BH hypothesis was 
first put forward by Ryan~\cite{ry}, who considered a general 
stationary, axisymmetric, asymptotically flat, vacuum space-time. 
Such a generic space-time can be used to describe the 
gravitational field around a central object, whatever its nature, 
and its metric can be expressed in terms of the mass moments 
$M_\ell$ and current moments $S_\ell$~\cite{mom}. Assuming 
reflection symmetry, the odd $M$-moments and even $S$-moments 
are identically zero, so that the non-vanishing moments are the 
mass $M_0=M$, the mass quadrupole $M_2=Q$ and the 
higher-order even terms $M_4, M_6,\ldots$, as well as the angular 
momentum $S_1=J$, the current octupole $S_3$ and the 
higher-order odd terms $S_5, S_7, \ldots$. In the case of a Kerr 
BH, all the moments $M_\ell$ and $S_\ell$ are locked to the mass 
and angular momentum by the following relation:
\begin{equation}\label{kerrMultipoles}
M_\ell+{\rm i}S_\ell=M\left({\rm i}\frac{J}{M} \right)^\ell \, .
\end{equation}
By measuring at least three non-trivial multiple moments of the
space-time around an astrophysical BH candidate (e.g. $M$, 
$J$, and $Q$), one over-constraints the theory and can test the
Kerr nature of the object.

A post-Newtonian approach with generic $M$, $J$, and $Q$ can
do the job only if we have the possibility of measuring with excellent
accuracy the geometry relatively far from the compact object,
where the contribution of higher-order moments is negligible.
The advantage of a post-Newtonian approach is that we do not
need specific assumptions about the space-time close to the BH.
The disadvantage is that very accurate measurements are
necessary, while they may be out of reach in realistic astrophysical 
situations.

The differences between a Kerr and a non-Kerr BH are more and 
more evident if we probe the space-time closer to the object, where 
gravity is strong. However, we have to consider a specific metric 
and take all the special and general relativistic effects into account. 
In this case, there are basically two approaches:
\begin{enumerate}
\item We assume that the space-time around the compact object can 
be described by an exact solution of the Einstein's vacuum equations 
which, because of the no-hair theorem, cannot be a BH. The idea 
on the basis of this choice is that this metric can correctly describe 
the exterior gravitational field around the compact object and holds 
up to the ``surface'' of the object. The latter should cover all the 
pathological features (e.g. space-time singularities and regions with 
close time-like curves) of the space-time. 
\item We consider a metric with a regular event horizon: this is a 
true BH but, as follows from the no-hair theorem, it is not a solution 
of the Einstein's vacuum equations. The advantage of this 
approach is that the position of the event horizon is well defined, 
while in the first approach there is still some arbitrariness to be fixed.
\end{enumerate}
In both cases, we eventually have a metric more general than the 
Kerr solution and that includes the Kerr space-time as special case.
So, we have a compact object characterized by a mass $M$, a spin 
angular momentum $J$, and one or more ``deformation parameters'',
say $\delta$ (or $\{\delta_i\}$ with $i=1,2,\ldots$), measuring (hopefully
generic) deviations from the Kerr geometry. When $\delta = 0$,
the object is a Kerr BH. One can then consider a particular astrophysical
phenomenon, compare the theoretical predictions in this general 
space-time with the observational data, and constrain the deformation
parameter $\delta$. If the observations demand $\delta = 0$, they
confirm the Kerr BH hypothesis.

The Tomimatsu-Sato space-time was the first non-Kerr solution of 
the Einstein's vacuum equations discovered~\cite{ts1,ts12,ts13,ts14}. 
It is stationary, 
axisymmetric, and asymptotically flat and it has three free parameters 
(mass, spin, and deformation parameter). However, the metric is quite 
complicated, there is no analytic form for any value of the deformation 
parameter, and deviations from the Kerr geometry are not very generic
(for instance, the Kerr solution is recovered even for $|a_*| = 1$, 
independently of the value of the deformation parameter). A metric 
more useful to test the Kerr BH hypothesis is the Manko-Novikov 
solution~\cite{mn,mn2}. It is still a stationary, axisymmetric, and asymptotically 
flat exact solution of the Einstein's vacuum equations, but it has an 
infinite number of free parameters: here the object has a mass $M$, a 
spin angular momentum $J$, and arbitrary mass multipole moments
$M_\ell$ ($\ell=1,2,\ldots$), 
while the current multipole moments $S_\ell$ of order higher 
than 1 are fixed by the mass multipole moments. Some authors have 
also proposed approximated solutions of the Einstein's vacuum 
equations: these metrics are solutions in GR only up to some order of 
one or more expansion parameters. This is the case, for instance, of 
the ``quasi-Kerr'' metric proposed by Glampedakis and Babak~\cite{babak} 
and of the ``bumpy BHs'' introduced by the MIT group~\cite{mit1,mit12,mit13}.

In order to be able to probe the geometry of the space-time very close 
to a BH candidate and study very rapidly-rotating objects (for which
possible deviations from the Kerr metric should produce stronger 
effects, see next section), it may be more convenient to use a metric
describing a BH, without the ambiguity related to the position of the 
``surface'' of the object. These space-time can be seen as BHs in 
alternative theories of gravity, even if we do not know the gravity theory
they would belong to. The first theoretical framework in this direction 
was proposed in~\cite{mit2}: these objects are called ``bumpy BHs in 
alternative theories of gravity''. Another family of metrics describing 
non-GR BHs was introduced by Johannsen and Psaltis in Ref.~\cite{ps}.


\section{Properties of non-Kerr black holes \label{s-3}}

If the astrophysical BH candidates are not Kerr BHs, some of their
fundamental properties are likely very different from the ones 
predicted by GR. If we know some generic features of non-Kerr BHs, 
we can hopefully figure out generic observational signatures to 
test the nature of the astrophysical candidates.

\subsection{Spin parameter}

A fundamental limit for a Kerr BH is the bound $|a_*| \le 1$. This 
is just the condition for the existence of the event horizon. In absence
of a horizon there would be a naked singularity, which is forbidden 
by the weak cosmic censorship conjecture~\cite{wccc}. Even if 
it is not yet clear if naked singularities can be created in Nature,
and therefore if the weak cosmic censorship conjecture can be
violated (see e.g.~\cite{pankaj,pankaj2} and references therein), it is
apparently impossible to make a star collapse with $|a_*| > 
1$~\cite{wccc2,wccc3} or to overspin an already existing BH to
$|a_*| > 1$~\cite{e2}. Moreover, even if it were possible to create
a Kerr naked singularity, the space-time would be unstable and
it should quickly decay~\cite{dotti}. The same would be true if the
singularity were replaced by a very compact object arising from
new physics, independently of its nature, because of the ergoregion
instability~\cite{e1}.

The story changes if the compact object is not a Kerr BH: in this 
case, the maximum value of $|a_*|$ may be either larger or smaller 
than 1, depending on the geometry of the space-time around
the compact object, on the exact nature of the compact object, and
on the gravity theory. Generally speaking, objects with $|a_*| > 1$
are quite common in the Universe and even the value of the spin 
parameter of the Earth is about $10^3$. However, a priori it is not
obvious if it is possible to create even a compact object with 
$|a_*| > 1$.

In Refs.~\cite{spin1,spin12,spin13}, 
I considered compact objects characterized by
three free parameters (mass $M$, spin parameter $a_*$, and
quadrupole moment $Q$) and I studied the evolution of $a_*$ as
a consequence of the accretion process from a thin disk. It is
indeed well known that accretion from a thin disk is a quite efficient
mechanism to spin-up a compact body. In the standard theory of
thin disks (Novikov-Thorne model~\cite{n-t-m}), the disk is on the equatorial
plane, the gas particles move on nearly geodesic circular orbits,
and the inner radius of the disk is at the innermost stable circular
orbit (ISCO). When the gas reaches the ISCO, it quickly plunges into
the BH and crosses the event horizon, with negligible emission of 
additional radiation. So, the BH changes its mass by $\delta M = 
E_{\rm ISCO} \delta m$ and its spin by $\delta J = L_{\rm ISCO} \delta 
m$, where $E_{\rm ISCO}$ and $L_{\rm ISCO}$ are respectively the
specific energy and the $z$-component of the specific angular momentum 
of a particle at the ISCO radius, while $\delta m$ is the gas rest-mass.
When the Novikov-Thorne model can be applied, the evolution of 
the spin parameter is thus governed by the following equation:
\begin{eqnarray}\label{e-spin}
\frac{d a_*}{d \ln M} = \frac{1}{M}
\frac{L_{\rm ISCO}}{E_{\rm ISCO}}  - 2 a_* \, .
\end{eqnarray}
Recent numerical simulations support this model for thin disks around
Kerr BHs~\cite{cfa,cfa2} (see however Ref.~\cite{kro}). The equilibrium
spin parameter is reached when the right hand side of Eq.~(\ref{e-spin})
is zero. For a Kerr BH, this occurs when $a_*^{eq} = 1$. Including
the small effect of the radiation emitted by the disk and captured by 
the BH, one finds the famous Thorne's limit $a_*^{eq} = 0.998$~\cite{th}.
On the other hand, if the compact object is more oblate than a
Kerr BH, the equilibrium spin parameter is larger than 1, and its value
increases as the object becomes more and more oblate. The situation
is more complicated for objects more prolate than a Kerr BH, and the
value of the equilibrium spin parameter can be either larger or smaller
than 1. The result of Refs.~\cite{spin1,spin12,spin13} is thus that compact objects
with $|a_*| > 1$ can be created, at least in principle. Depending on 
the exact nature of the compact object and on the gravity theory, 
there is still the possibility that the equilibrium spin parameter
predicted by Eq.~(\ref{e-spin}) can never be reached, because the 
compact object becomes unstable at a lower value of $|a_*|$. 
For instance, a similar situation occurs for neutron stars: the
accretion process can spin them up to a frequency $\sim 1$~kHz,
but the existence of unstable modes prevents these objects from 
rotating at higher frequencies~\cite{neutron}.

\subsection{Event horizon \label{ss-eh}}

The event horizon of a BH is defined as the boundary of the causal 
past of future null infinity. The spatial topology of the event horizon 
at a given time is the intersection of the Cauchy hypersurface at that 
time with the event horizon. In the Kerr space-time, the spatial
topology of the event horizon is a 2-sphere. However, the topology does
not change even if the space-time is not exactly described by the Kerr
geometry, e.g. like in the case of a BH surrounded by a disk of accretion.
The Hawking's theorem indeed ensures that, in 4-dimensional 
GR, the spatial topology of the event horizon must be always a 
2-sphere in the stationary case, under the main assumptions of 
asymptotically flat space-time and validity of the dominant energy 
condition~\cite{hawking}. In the non-stationary case, BHs with a 
toroidal spatial topology can form, but the hole must quickly close 
up, before a light ray can pass through~\cite{tedj}. Numerical 
simulations confirm the theoretical results~\cite{teuk,teuk2}.

Non-Kerr BHs may instead have topologically non-trivial event 
horizons, as they are not solutions of the Einstein's equations. 
In Ref.~\cite{topo}, I have even argued that topologically non-trivial
event horizons may be a quite common feature for rapidly-rotating
non-Kerr BHs, providing two explicit examples in which the horizon of the
BH changes topology above a critical value of the spin parameter.
The basic mechanism is the following. A Kerr BH has an outer 
horizon of radius $r_+$ and an inner horizon of radius $r_-$. 
As $|a_*|$ increases, $r_+$ decreases, while $r_-$ increases. 
For $|a_*| = 1$, there is only one horizon ($r_+ = r_-$) and, for 
$|a_*| > 1$, there is no horizon. In general, however, the outer and 
the inner horizons may not have the same shape. If this is the
case, when $|a_*|$ increases, the two horizons approach each 
other, but eventually they merge together forming a single horizon 
with non-trivial topology. Interestingly, such rapidly-rotating BHs can 
be easily created. Indeed, the topology transition can be potentially 
induced by the accretion of material from a thin disk, which 
should be a quite common event in the Universe.

\subsection{Accretion process \label{ss-ap}}

The geometry of the space-time around astrophysical BH 
candidates can be probed by studying the properties of the
electromagnetic radiation emitted by the gas of accretion. It is 
thus important to figure out clearly the accretion process itself 
and have under control all the astrophysical processes.

If the gas around the compact object has negligible angular momentum,
a Bondi-like accretion can correctly describe the evolution of the
system. For example, this may be the case of a BH
accreting from the interstellar medium or of one belonging to a binary
system in which the companion is massive and has a strong stellar 
wind. Here, the accretion process is basically determined by the balance 
between the gravitational force, which attracts the gas towards the
central body, and the gas pressure, which increases as the gas becomes
more and more compressed and hampers the accretion process.
If the gravitational force is stronger, the compact object can swallow 
a larger amount of matter without problems. If the gravitational force 
is weaker, the accretion process is more difficult and the production 
of outflows is favored. Around naked singularities, the gravitational 
force is typically much weaker (indeed, unlike a BH, a naked 
singularity cannot trap
light rays) and can be even repulsive. For more details, see 
Refs.~\cite{super,super2,super3,super4,super5,super6}.

The formation of an accretion disk around the compact object is 
possible when the gas around the body has significant angular 
momentum. The disk is geometrically thin and optically thick 
when the gravitational energy of the falling gas can be efficiently 
radiated away. Thin disks are described by the Novikov-Thorne model. 
One of the key ingredients is the existence of an ISCO: circular orbits inside 
the ISCO are radially unstable and therefore, as the gas 
reaches the ISCO, it quickly plunges onto the compact object and 
crosses the event horizon, without significant emission of  additional 
radiation. The evolution of the spin parameter of the BH is given by 
Eq.~(\ref{e-spin}) and the radiative efficiency $\eta$, defined by 
$L_{\rm acc} = \eta \dot{M}$, where $L_{\rm acc}$ is the luminosity 
due to the accretion process and $\dot{M}$ is the mass accretion 
rate, is simply $\eta = 1 - E_{\rm ISCO}$.

If the geometry around the compact object is not described by the
Kerr metric, one can find even other scenarios~\cite{bb2}.
In particular, if the object is more prolate than a Kerr BH, circular
orbits on the equatorial plane may be even vertically unstable
(in the Kerr background, all the circular orbits on the equatorial
plane are vertically stable) and, in addition to an outer region
with stable circular orbits delimited by the ISCO, one may find 
regions with stable circular orbits even closer to the compact object. 
At least for the subclass of Manko-Novikov space-times studied 
in~\cite{bb2}, there are four qualitatively different final stages of 
accretion:
\begin{enumerate}
\item The ISCO is {\it radially} unstable, and the gas plunges 
into the compact object remaining roughly on the equatorial 
plane and without emitting significant radiation. 
This is the same scenario as in the Kerr case.
\item The ISCO is {\it radially} unstable and the gas 
plunges, but does not reach the compact object. Instead, 
it gets trapped between the object and the ISCO,  
forming a thick disk.
\item The ISCO is {\it vertically} unstable, and the gas plunges 
into the compact object {\it outside} the equatorial plane and 
without emitting significant radiation.
\item The ISCO is {\it vertically} unstable and the gas 
plunges, but does not reach the compact object. Instead,
it gets trapped between the object and the ISCO and        
forms two thick disks, above and below the equatorial plane.
\end{enumerate}
The scenarios (2) and (4) occur only in a limited range of the 
parameters of these space-times. Nevertheless, they have
quite peculiar features and may be likely tested with
future observations. Because of the presence of a thick disk
inside the ISCO, the evolution of $a_*$ is given by 
Eq.~(\ref{e-spin}) with $E_{\rm inner}$ and $L_{\rm inner}$ 
replacing respectively $E_{\rm ISCO}$ and $L_{\rm ISCO}$,
where $E_{\rm inner}$ and $L_{\rm inner}$ are the
specific energy and the $z$-component of the 
specific angular momentum at the
inner edge of the thick disk, and the radiative efficiency
becomes $\eta = 1 - E_{\rm inner}$.


\section{Observational tests \label{s-4}}

Generally speaking, the Kerr BH hypothesis can be tested by
using the same techniques through which astronomers can
measure the spin parameter of a BH, assuming the geometry
of the space-time is described by the Kerr metric. In what
follows, I discuss in some details the three approaches of which
I have some experience, while other possibilities are just 
mentioned in the last subsection.

\subsection{Radiative efficiency}

The radiative efficiency $\eta$ is defined by the relation 
$L_{\rm acc} = \eta \dot{M}$, where $L_{\rm acc}$ is the accretion 
luminosity and $\dot{M}$ is the mass accretion rate. In the case
of a BH, $\eta$ may be even extremely small, because the gas
can cross the event horizon before it can radiate away the
energy of the gravitational potential. For example, in a spherically
symmetric and adiabatic accretion onto a Schwarzschild BH
(Bondi accretion), $\eta \sim 10^{-4}$. High values of the radiative
efficiency can be easily obtained in presence of a thin accretion disk,
where $\eta = 1 - E_{\rm ISCO}$\footnote{As discussed in 
Subsec.~\ref{ss-ap}, $\eta = 1 - E_{\rm inner}$ in those special
cases in which the gas forms a thick disk between the ISCO and 
the BH.}; for a Schwarzschild BH ($a_* = 0$), we
find $\eta = 0.057$, while for a maximally-rotating Kerr BH
and a corotating disk ($a_* = 1$), $\eta = 0.42$.
When the compact
object has a solid surface, the picture is instead much more 
complicated, as the gas may radiate additional energy 
when it hits the surface of the body; anyway, this does not seem 
the case for the BH candidates~\cite{horizon,horizon2}.

In general, it is difficult to estimate the radiative efficiency of a
BH candidate, because it is not possible to measure the mass 
accretion rate. However, one can estimate the mean radiative
efficiency of active galactic nuclei (AGN) by using the Soltan's
argument~\cite{soltan}, which relates the mean energy density in 
the contemporary Universe radiated by the super-massive BHs 
with the today mean mass density of these objects. In the final 
result, there are definitively several sources of uncertainty.
However, a reliable lower bound is thought to be $\eta > 
0.15$~\cite{elvis,elvis2}, especially if one restrict the attention to the 
most massive objects.

For a Kerr BH, since the radiative efficiency is $\eta = 1 - E_{\rm ISCO}$ 
and increases as $a_*$ increases, $\eta > 0.15$ corresponds
to $a_* > 0.89$. If we do not assume that the super-massive
objects in galactic nuclei are Kerr BHs, it is possible to get a constraint
on the mean deformation parameter of AGN. In Ref.~\cite{eta}, I considered a
subclass of the Manko-Novikov space-times, in which the deformation
parameter was the quadrupole moment $Q$ of the object. Defining the
dimensionless parameter $q$ as $Q = - (1 + q) a_*^2 M^3$ (for 
$q = 0$, we recover exactly the Kerr solution), observations require:
\begin{eqnarray}\label{e-q}
-2.01 < q < 0.14 \, ,
\end{eqnarray}
Let us notice that, in the case of a self-gravitating fluid like a neutron 
star, one would expect $q > 1$. In other words, BH candidates are
much stiffer than ordinary matter.

Since the most efficient way to spin a compact object up is
through the accretion process from a thin disk,
the maximum possible value of the spin parameter of a super-massive
object at the center of a galaxy is given by the equilibrium spin
parameter $a_*^{eq}$ of Eq.~(\ref{e-spin}). The exact value depends
on the deformation parameter. If we require that a BH candidate
must be able to have $\eta > 0.15$ with $a_* \le a_*^{eq}$, we find 
the maximum value of the spin parameter for the super-massive
objects in galactic nuclei~\cite{bspin,bspin2}:
\begin{eqnarray}\label{e-a}
|a_*| \lesssim 1.2 \, ,
\end{eqnarray}
which is basically independent of the choice of the theoretical
framework. This argument cannot be applied to the stellar-mass
BH candidates because the value of the spin parameter of the
latter reflects the one at the time of their formation;
that is, it is determined by more complex physics involving the
gravitational collapse.

As shown in this subsection, the estimate of the radiative efficiency
can already be used to provide interesting constraints on the nature
of the BH candidates. However, the constraints are
weak and presumably they cannot be significantly improved in the 
near future, as there are several sources of uncertainty. On the other
hand, the approaches discussed in the next subsections are more 
complicated, but much more promising to get stronger and more
robust bounds.

\subsection{Continuum fitting method}

The X-ray spectrum of stellar-mass BH candidates has often a
soft component ($< 10$~keV), which is thought to be the thermal
spectrum of a geometrically thin and optically thick disk of
accretion. In the Novikov-Thorne model, the observed spectrum 
of a thin disk around a BH depends only on the background 
metric, the mass accretion rate, the distance of the observer, and 
the viewing angle. Assuming the Kerr background, one can 
measure the spin parameter from the observational data. This 
technique is called continuum-fitting method and at present it has 
been used to estimate the spin of a few stellar-mass BH 
candidates~\cite{cfm,cfm2}\footnote{For super-massive BHs, the disk 
temperature is lower (the effective temperature scales like 
$M^{-0.25}$) and this approach cannot be applied.}. 
Basically, one has to get independent measurements of the the 
mass of the object, its distance from us, and the inclination angle 
of the disk, and then it is possible to fit the soft X-ray component 
of the source and deduce $a_*$ and $\dot{M}$. The key-point is
that there is a one to one correspondence between the value
of $a_*$ and the one of the radiative efficiency $\eta$.

Relaxing the Kerr BH hypothesis, one can probe the geometry 
around the stellar-mass BH candidates~\cite{bb1} (see also 
Refs.~\cite{cfm-s,cfm-s2,cfm-s3,cfm-s4,cfm-s5,cfm-s6,cfm-s7} 
for tests of more specific models). Generally
speaking, current X-ray data are not so good to break the 
degeneracy between the spin parameter and the deformation 
parameter, but it is only possible to constrain a combination of
the two. However, the thermal spectrum of a thin disk around
a rapidly-rotating Kerr BH can be hardly mimicked by a compact
object very different from a Kerr BH. So, if we observe a spectrum
that seems to be generated around a very rapidly-rotating Kerr
BH, we can constrain the deformation parameter, independently of the
value of its spin. In principle, we could 
also discover deviations from the Kerr geometry if we find that
the thermal spectrum of the disk is too hard even for a Kerr
BH with $a_* = 1$. It may be interesting to test this possibility 
with the spectrum of the high-spin BH candidate 
GRS1915+105~\cite{grs}, when future more accurate measurements 
of the distance to this object will be available. However, 
significant work has still to be done, especially in the case of
fast-rotating objects, before using the continuum fitting method to
probe the geometry around the BH candidates~\cite{bb1}.

\subsection{Direct imaging}

The capability of very long baseline interferometry (VLBI) has
improved significantly at short wavelength and it is now widely believed 
that within 5-10 years it will be possible to observe the direct image 
of the accretion flow around nearby super-massive BH candidates with 
a resolution comparable to their event horizon~\cite{doel,doel2}. If the disk
is optically thin (which is always the case for sufficiently high frequencies)
and geometrically thick, it will be possible to observe the BH
``shadow'', i.e. a dark area over a bright background~\cite{sha0,sha02}. 
While the intensity
map of the image depends on the details of the accretion process, the
contour of the shadow is determined exclusively by the geometry of the
space-time around the compact object. The observation of the shadow
can thus be used to test the Kerr BH hypothesis, as first suggested 
in~\cite{sha1,sha12,sha13} and further explored in~\cite{sha2,sha22}.

The contour of the shadow is the photon capture surface as seen by
a distant observer. As light rays are bent by the gravitational field of
a massive object, the size of the shadow is always larger than the one of the
photon capture surface. In the case of rotating objects, in general
the shadow is not symmetric with respect to the axis of the spin of the 
BH, because the capture radius for corotating photons is
smaller than the one for counterrotating ones; the effect is maximum 
for an observer on the equatorial plane and goes to zero for one
along the $z$-axis. For generic space-times, the contour of the
shadow is computed by considering the photons crossing perpendicularly
the image plane of the distant observer and integrating numerically 
backward in time the geodesic equations. All the points on the image 
plane of the observer whose trajectories cross the BH horizon make 
the shadow.

As discussed in Subsec.~\ref{ss-eh}, rapidly-rotating non-GR BHs
may have a topologically non-trivial event horizon. One can see
that in these space-times the central singularity is naked. In other
words, in these scenarios it is easy to overspin a BH and violate
the weak cosmic censorship conjecture. If such rapidly-rotating
BHs exist, a distant observer may see also the central singularity or,
more likely, the quantum gravity region replacing the central 
singularity~\cite{topo2}.

\subsection{Other tests}

In addition to the continuum fitting method, another famous
technique among astronomers to measure the spin parameter
of BH candidates is the approach of the relativistic iron line~\cite{reyn}.
Basically, one sees a broad spectral line which is interpreted as
fluorescent iron K$\alpha$ emission from cool gas in the accretion
disk. The rest energy of the iron line is 6.4~keV, while the 
observed line is broad because it is affected by Doppler boosting,
frame dragging, and gravitational redshift. 
The advantage with the continuum fitting 
method is that this technique can be applied either to stellar-mass
and super-massive BH candidates. However, the physics is more 
complicated, we have no way to predict {\it a priori} the 
intrinsic surface brightness profile of the K$\alpha$ line, 
and even the basic model is subject to critiques~\cite{tita}.
A preliminary study to use the K$\alpha$ line to test the Kerr BH
hypothesis is reported in Ref.~\cite{iron}.

The geometry of the space-time around BH candidates can also 
be mapped by studying the orbital motion of individual stars.
The latter are massive objects, so they move along the geodesics
of the space-time and, unlike the gas particles, their motion is
not affected by the presence of electromagnetic fields. However,
they are usually relatively far from the BH candidate, where
gravity is weak, and therefore very accurate measurements
are necessary. In the literature, there are two proposals in this
direction. The study of the motion of a radio pulsar around a stellar-mass
BH companion~\cite{wex}; for the time being, however, no BH-pulsar
binary system is known.
Astrometric monitoring of stars orbiting at mpc distances from
SgrA$^\star$~\cite{will2,will22}; in principle, it would indeed be possible to measure 
the spin $J$ and the mass-quadrupole moment $Q$ of the 
super-massive BH candidate at the center of the Galaxy.
While future infrared experiments could be able to observe such short 
period stars, it is not clear if the idea can work, because the presence 
of unknown and unseen objects closer to the BH candidate would 
also affect the motion of these stars and spoil the measurement.

Another interesting proposal was put forward in Ref.~\cite{bhb}
and uses the available optical data of the BL Lacertae object OJ287.
The system is modeled as a spinning primary BH with an accretion
disk and a non-spinning less massive secondary BH. If we 
assume that the observed outbursts from 1913 to 2007 arise when
the secondary BH crosses the accretion disk and we fit the data,
we can constrain the spin and the quadrupole moment of the
primary BH.

Future gravitational wave astronomy may open new possibilities
and test the nature of the BH candidates with excellent accuracy.
That should be possible in at least two ways:

{\it Observations of extreme-mass ratio inspirals (EMRIs)}. These
are systems consisting of a stellar-mass compact object orbiting
a super-massive BH candidate. Since future space-based
gravitational waves antennas like LISA (or a similar ESA-led mission)
will be able to follow the stellar-mass object for millions of orbits
around the super-massive BH candidate, the space-time around 
the latter can be mapped with very high accuracy. Any deviation
from the Kerr geometry will lead to a phase difference in the
gravitational waveforms that grows with the number of observed
cycles. The technique is very promising, it is not very sensitive to
the exact field equations of the gravity theory, and it has been
studied in details by many authors~\cite{babak,emris,emris2,emris3,emris4,emris5,emris6}.

{\it Observations of quasi-normal modes (QNMs)}.
Ground-based detectors like the Einstein Telescope will be
able to observe the QNMs of stellar-mass objects, while the ones
of super-massive BH candidates will require space-based missions
like LISA. For a Kerr BH, the frequencies of these modes depend
only on its mass $M$ and spin $J$. The identification of at least
three modes can thus be used to test the nature of a BH candidate.
As the QNMs are more sensitive to the specific field equations of 
the gravity theory, the interpretation of the data is more complicated 
with respect to the first approach of the EMRIs. 
For more details, see Refs.~\cite{qnms,qnms2,qnms3,qnms4}.


\section{Conclusions \label{s-5}}

Up to now, GR has successfully passed all experimental tests, but the
theory has never been checked under extreme conditions, where new
physics can more likely appear. In particular, we currently believe that 
the final product of the gravitational collapse is a Kerr BH, but
we have no observational confirmations. Astronomers have
already discovered several good astrophysical candidates
and recently there has been an increasing interest in the
possibility of testing the Kerr nature of these objects with present
and future experiments. 
If the BH candidates are not the BH predicted by GR, they
likely have different fundamental properties: for instance, they
may violate the bound $|a_*| \le 1$, they may have a
topologically non-trivial event horizon, and even the accretion
process may present peculiar features.
For the time being, one can probe
the geometry around a BH candidate by using the techniques
developed to estimate the spin parameter of these objects
under the assumption they are Kerr BHs; the continuum fitting
method and the study of relativistic lines are the two most
popular approaches. With the already available X-ray data,
we can test the Kerr BH hypothesis, at least when all the astrophysical 
processes are well understood. In the future, we can hope
to get more reliable and precise measurements with the
advent of new observational facilities: VLBI arrays will 
be able to observe the shadow of nearby super-massive BHs,
thus detecting the apparent photon capture surface of the
space-time,
ground-based gravitational wave detectors like the Einstein Telescope
will be able to test the Kerr BH hypothesis by detecting the QNMs of
stellar-mass BHs, and space-based
gravitational wave antennas like LISA or a similar ESA-led mission
will be able to observe the gravitational waves emitted by EMRIs 
and test the nature of super-massive BHs.


\section*{Acknowledgments}

I would like to thank my collaborators in the works reviewed in this
paper: Enrico Barausse, Francesco Caravelli, Katie Freese, Tomohiro 
Harada, Leonardo Modesto, Rohta Takahashi, and Naoki Yoshida.
I am grateful to Federico Urban, for reading a preliminary version 
of the manuscript and providing useful feedback, and Luciano
Rezzolla, for useful comments about the weak cosmic censorship
conjecture. This work was supported by Humboldt Foundation.

\end{document}